# Prospects for endurance augmentation of small unmanned systems using butane-fueled thermoelectric generation

Morgan Williamson[1,*], Aditya Rao[2], Evan Segura[2], Bryson Wylie[2] and Matthew Hall[2]

[1] Applied Research Laboratories, University of Texas at Austin, Austin, TX 78712, USA
[2] Department of Mechanical Engineering, University of Texas at Austin, Austin, TX 78712, USA
* Correspondence: morgan.williamson@arlut.utexas.edu

**Abstract:** We investigate the potential of enhancing small (<20 kg) drone endurance by exploiting the high energy density of hydrocarbons using a prototype generator based on commercial-off-the-shelf (COTS) thermoelectric energy conversion technology. A proof-of-concept prototype was developed to vet design and engineering challenges and to bolster validity of resultant conclusions. The combination of the prototype performance and modeling suggests that endurance augmentation remains a difficult technical challenge with no clear immediate remedy despite many expectant alternatives. Across a sample of representative drones including ground- and air-based, multicopter and fixed wing drones, we report the following: from their current maximum values of 12%, thermoelectric (TE) generator module efficiencies must increase by over two times to achieve endurance parity with lithium batteries for VTOL multicopters. On the other hand, current TE efficiencies can compete with lithium batteries for some low power fixed wing and ground-based drones. Technical contributors for these results include weight of non-energy contributing components, low specific power and the associated tradeoff between specific power and specific energy due to fuel mass fraction, and lastly, low efficiencies.

## 1. Introduction

Recent geopolitical conflicts have punctuated the strategic importance of drones of various types on the battlefield. Small (group 1) drones employed as intelligence, surveillance, and reconnaissance (ISR) and ordinance delivery platforms have proved to be particularly poignant. Furthermore, the trend of increasing importance of low-cost attritable resources is not showing signs of abating. Such development is couched within the backdrop of small drones' steady rise in utility and popularity for myriad uses including those outside of defense.

One particularly vexing weakness of small drones (especially multicopters) happens to be their limited endurance. In general, a high performing small multicopter drone able to carry a minimally useful payload is able to stay aloft for around 10-30 minutes. Needless to say, this constraint subsequently has a large effect on real world range and utility. To some extent the endurance weakness may be alleviated by employing fixed wing small drones instead, however, often at the cost of sacrificing hovering maneuverability and the flexibility associated with takeoff and landing that a VTOL capability affords.

Drones above group 1 typically do not possess as grave an endurance weakness, which is the result of multiple factors. First, scaling laws tend to work in the favor of larger electric drones. Case in point, very small so-called tiny whoops only achieve 3 to

5-minute flight times. At the other end of the scale, saturating the practical endurance of battery technology today, the full-size training aircraft Pipistrel Velis Electro attains a maximum flight endurance of 55 minutes. Second, larger drones are typically able to exploit internal combustion engines (ICEs) for thrust, which also scale unfavorably to smaller sizes due to their surface area to volume ratio. Therefore, there exists a nominal threshold for aircraft mass around the 20 kg mark in which an ICE's weight can be justified, and which in turn affords 10 to 24-hour endurance. Therefore, the stepwise jump in performance from group 1 to group 2 constitutes over an order of magnitude jump in endurance coinciding with the ability to utilize internal combustion engines and hydrocarbon fuels.

The unfortunate gap in endurance between group 1 electric UASs and group 2 ICE-power UASs has persisted despite earnest attempts at address, most promising and mobilized of which may be the extensive field of battery research. Indeed, there is no shortage of innovative developments in battery research that present potential solutions. However, the technology readiness level of such developments often takes decades to breach the commercial market. For instance, the research-to-commercialization cycle of lithium chemistry spanned three decades [1]. As a result, a point of intentional importance for this work is the contemporaneous fieldability of any alternative solution.

Alternative energy sources have been successfully implemented at the prototype level for small drones such as integration of hydrogen fuel cell technology.[1] While fuel cells show promise, the high pressures required for storing the hydrogen fuel constitute an explosive risk and the storage tanks tend to be relatively heavy. The use of fuel cell technology could also entail greater complications in fuel delivery logistics and potentially insufficient specific power and power density. Although, research into breaking free of the use of hydrogen and using more pedestrian fuel chemistries is developing [2, 3].

In this work, we investigate the prospects for increasing small drone endurance using currently available COTS solutions employing thermoelectric generation modules powered by readily available backpacking butane heat sources and accessories. Thermoelectrics were chosen for their simplicity, potentially silent operation, lack of moving parts, and their generality in interfacing with virtually any heat source [4–8]. To wit, radioisotope thermoelectric generators have proved immensely effective for space exploration missions [9]. Butane (and/or iso-butane) was chosen as a fuel source due to its high energy density and gaseous state at standard temperature and pressure, which simplifies internal fuel delivery obviating the need for any pump. Butane also possesses a low vapor pressure of 2.5 bar, which enables the use of thin-walled, lightweight fuel tanks, crucial for any airborne application. Also, due to the prevalence of backpack hiking, butane and iso-butane fuel delivery logistics are well established and afford economies of scale for any potential practical implementation of such an approach. The guiding intention with the thermoelectric approach is to breach the performance gap of unmanned systems between groups 1 and 2 by converting the high energy density of hydrocarbons into a power source compatible with electric group 1 drones.

The current dominant battery technology for small drones consists of lithium-polymer and lithium-ion chemistries. Typical values for the specific energies for each are: 150 Wh/kg and 200 Wh/kg, respectively. Lithium-ion batteries are commonly used for longer endurance applications with the caveat that the specific power of lithium-polymer batteries (around 3500 W/kg) will not be attainable. On the other hand, butane has a specific energy of 13,600 Wh/kg, a value similar to other hydrocarbons.

---

[1] e.g. Hylium Industries, Inc. HyliumX

The shear amount of energy stored in hydrocarbons suggests that if there was a method to capture the energy content, despite the inefficiencies of conversion, it may have a tantalizing chance of exceeding the specific energy of current battery technology, the salient metric presently limiting drone endurance. Case in point, despite ~20% thermal efficiency, ICE-powered automobiles are able to reach over ten times the amount of range compared to electric vehicles, leading to the so-called "range anxiety" issue. This specific energy limitation of batteries forces the tendency for electric cars to possess a higher battery mass fractions than the fuel mass fractions of ICE-powered cars.

A back of the envelope calculation can determine the sufficient generalized "specific efficiency" for achieving parity with current battery technology.

$$150 \frac{Wh}{kg} = 13{,}600 \frac{Wh}{kg} \eta_s \tag{1}$$

By comparing the specific energies, such a figure, $\eta_s$, would be 1.5%, constituting the product of the fuel mass fraction, device efficiency, and exhaust loss efficiency sufficient to achieve parity.

The lowness of this figure motivates the investigation to seek out if thermoelectric generator systems can exceed the specific energy of lithium-based power systems.

$$e_{sys} = e_{fuel} \, \eta_s = e_{fuel} \frac{m_{fuel}}{m_{gen} + m_{fuel}} \eta_{dev} \, \eta_{exh} \tag{2}$$

Where $e_{sys}$ is the generator system specific energy, $e_{fuel}$ is the specific energy of the fuel, $m_{fuel}$ is the fuel mass, $m_{gen}$ is the generator mass, both of which combine to compose the fuel mass fraction. $\eta_{dev}$ is the device efficiency, and $\eta_{exh}$ is the efficiency due to exhaust loss. Regarding the energy efficiency estimates, the firm MicroPower offers COTS TE modules with maximum efficiencies of 12%, while typical estimates for exhaust loss efficiencies are 40% [5]. These figures suggest a minimum viable fuel mass ratio of 23% for a high-performance generator system to achieve parity with battery energy density.

In order to investigate the achievability of such a fuel mass fraction with adequate validity we designed and constructed a prototype. Physical construction was chosen over pure modeling due to the quality physical development has in eliciting wholly unforeseen challenges requiring resourceful solutions. The physical prototype would inform optimal sizing of components for sufficient power generation and provide a solid foundation for extrapolating potential performance through subsequent modeling.

In light of this backdrop, this work endeavors to answer the following questions: What level of small drone endurance is achievable today with COTS thermoelectric solutions? What are the limiting factors? And how much change in component performance would be required to achieve an appropriate advantage over current battery specific energy sufficient to warrant power system transition?

## 2. Materials and Methods

*2.1. Lightweight Prototype Thermoelectric Generator*

The apparatus in its simplest conceptual form consists of a fuel source, a burner, a thermoelectric module, heat sinks for the hot and cold sides of the TE module, and a method for supplying airflow to cool the cold side heat sink.

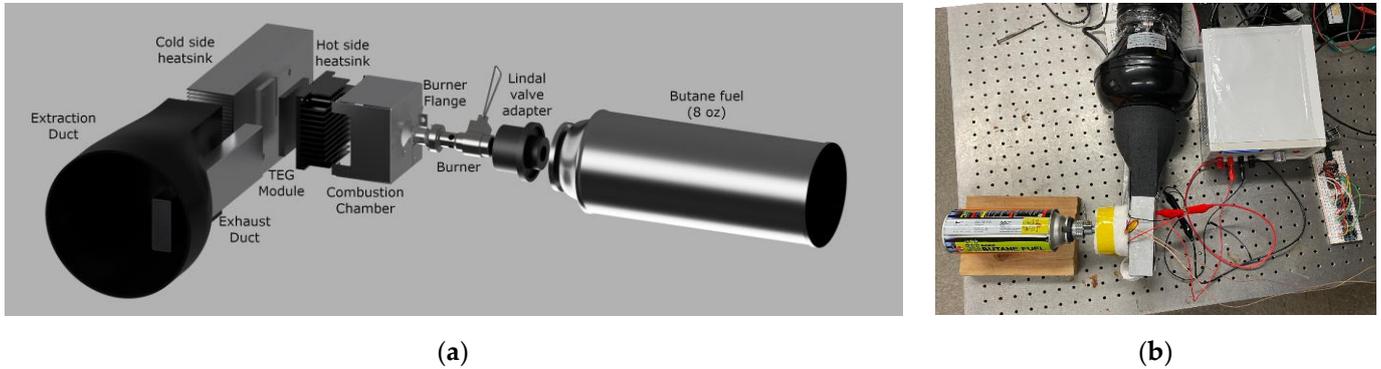

(**a**)        (**b**)

**Figure 1.** Depictions of the lightweight prototype thermoelectric generator. (**a**) An exploded view of the prototype 3D CAD model with labeled components; (**b**) Photo of the completed prototype in testing configuration with electronic load tester leads and thermocouples attached.

The following specific components were implemented to compose the apparatus: an 8 oz canister of butane with a bayonet mount supplies fuel through a bayonet-to-Lindal-valve adapter to a modified BRS-3000T burner. In this work we target an 800-watt thermal power production; the BRS-3000T's total output is capable of producing 2600 watts. The burner is housed in a combustion chamber fabricated using folded 24-gauge sheet aluminum surrounded by 1-inch-thick ceramic fiber insulation. The entry port for the burner is offset from center by approximately 1 cm in a direction away from the exhaust port in order to maximize the thermal dwell time. The burner flame impinges upon the 17 mm tall fins of a 50 mm square aluminum heat sink which spreads heat across the TE module. An exhaust port with a corresponding aluminum exhaust duct is positioned at the termination of the heat sink fins.

The TE module is a 40 mm square Bismuth Telluride example from Thermoelectric Conversion Systems Ltd. (TCS monTEG) capable of generating a stated maximum of 20 Watts at 5% efficiency and is sandwiched between the hot side and cold side heat sinks. A thin interface of Boron Nitride thermal paste is used on both sides of the module to maximize heat transfer. An adequate amount of thermal paste also served to ensure the thermocouples made sufficient contact with the TE module. The heat sinks are compressed together using titanium screws outfitted with Belleville washers to compensate for thermal expansion. Titanium was chosen as a high strength, lightweight heat brake to mitigate thermal short circuiting between the hot and cold sides due to its relatively low thermal conductivity. The four screws were torqued down to 1 Nm which corresponds to a 200 kg (50 kg for each screw) total compressive force in our tests[2], amply exceeding the required force of 90 kg for adequate thermal contact. The compressive method used in this work is unique since typical studies use a heavy external press to achieve the compressive load. The mobile or airborne applications in mind for this work necessitated verification of the viability of a weight optimized solution. For this same reason, liquid cooling was avoided as well, which has often been used in such studies [5].

---

[2] Tested using a button load cell read by a phidgets controller. Phidgets part descriptions and numbers: Button Load Cell - 200kg ID: 3137_0; Wheatstone Bridge Phidget ID: DAQ1500_0; VINT Hub Phidget ID: HUB0000_1.

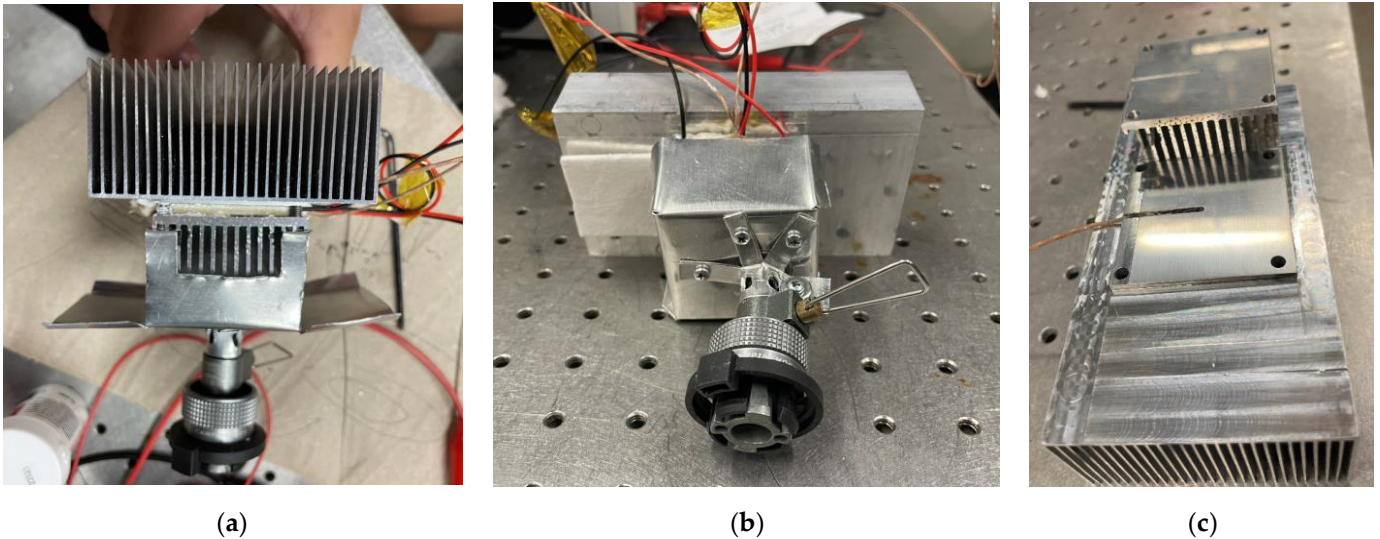

(a) (b) (c)

**Figure 2.** Detail views of the thermoelectric generator prototype. From bottom to top in (**a**) and (**b**): bayonet-to-Lindal-valve adapter, burner, burner attachment flange, combustion chamber (open in (**a**) during assembling), hot side heat sink, TE module, cold side heat sink; (**a**) View of the exhaust port immediately adjacent to the (smaller) hot side heat sink; (**b**) Detail view of closed combustion chamber and burner attachment flange; (**c**) Before assembly: Lapped surfaces of both cold and hot side heat sinks which interface through a thin layer of boron nitride with the TE module. Thermocouple channels and compression screw clearance holes visible.

The cold side heat sink is a modified extruded aluminum type with dimensions: 130x70x38 mm. The modifications consisted of removing unnecessary material totaling a weight savings of 82 grams (from 304 g), although more aggressive savings are attainable. (Itemized component masses are listed in Appendix A.) Also, the heat sink final sizing is determined by available airflow speed, which is dependent on application, thereby influencing weight further. (For multicopters, prop wash can be exploited for cooling flow (at the expense of some lift), while consistent forward airspeed would provide airflow for fixed wing drones.) The cold side and hot side heat sinks were lapped on a granite flat down to 15000 grit for optimal thermal contact. A 4-inch inline tube fan operating at 28 watts (5.5 m$^3$/min maximum flow rate) sucks ambient air into the intake of the cold side heat sink, which has a top cover plate to ensure channeled flow. At the output of the cold side heat sink a 3D printed rectangular-to-round duct transition channels air that has passed over the heat sink into a fume extraction duct. The combustion exhaust is entrained into the cooler flow from the cold side heat sink exit through a 24-gauge aluminum exhaust duct. In this work, the fume extraction apparatus is not counted as weight- or power-contributing components due to the reasons mentioned above regarding existing methods for airflow exploitation.

*2.2. Measurement Apparatus*

The measurement apparatus consists of a temperature logging subsystem and a power logging subsystem, each independent and controlled over USB using a computer. Three thermocouples (along with their amplifiers) were used to measure the hot and cold sides of the TE module as well as ambient temperature. Each heat sink possesses a 2 mm semicircular channel allowing thermocouple access to the hot and cold side center points of the TE module. These thermocouples are read out using an Arduino system that logs the temperature and timestamp of the hot and cold sides every second. The "Processing" language was used to log and save the data to comma separated value (CSV) files automatically for analysis. The power logging subsystem consisted of an electronic load tester (ZKETECH EBC-A10H) that was used to conduct power (specifically, current and voltage) measurements of the thermoelectric module using the ZKETECH software application every two seconds. A constant power discharge mode was used during tests requiring the use of a maximum power point tracker (MPPT) internal to the load tester. The most substantial portion of uncertainty affecting the power measurement regards the internal MPPT. Large swings in current and voltage could be seen during tests of the prototype likely due to instabilities in the MPPT tracking hardware or algorithm. As shown later in the results section figure 4, a sizable uncertainty is associated with the power measurements. We expect that this limitation also decreased observed power and energy production due to the quadratic dependence of power on voltage near the maximum power point.

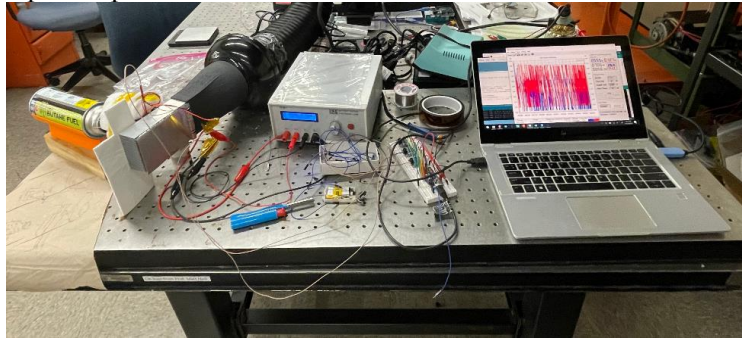

(**a**)

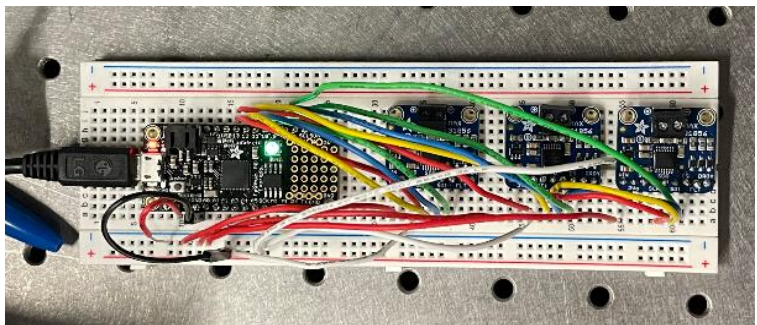

(**b**)

**Figure 3.** (**a**) The prototype generator under test on the left, the measurement apparatus to its right, complete with electronic load tester (white box, ZKETECH EBC-A10H), temperature logging circuit to its right, and monitoring PC on the far right; (**b**) Arduino circuit capturing temperature measurements. The main board on the left is an Adafruit ESP32-S3 Feather, while the three boards on the right are thermocouple amplifier boards, Adafruit Universal Thermocouple Amplifier MAX31856.

The prospect of remote and automated ignition of the combustion flame was explored using a modified arc generator circuit. Automated remote ignition enables the system itself to reignite the combustion flame if it would ever incidentally extinguish by intermittent lack of airflow, fuel interruption, etc. The ignition process could either be triggered manually or, for example, if the hot side temperature drops rapidly indicating lack of combustion. In a similar vein, a fuel control servo could also be installed in order to fully regulate the generation of power by the fuel flow rate through feedback control.

### 2.2. Testing Procedure

Testing the device consisted of recording temperature of both the hot and cold sides of the TE module as well as capturing the power generated by the device while in a combusting state. After burner ignition, there is a transient period of device heating when no power is being produced, yet fuel is being burned. Power production is required prior to the initiation of the data capture process within the electronic load tester. However, this complication is at least partially compensated by the cool-down process i.e. after fuel flow ceases there is a transient period in which power is still being produced. It should be noted that this transient behavior of TE systems is not encountered for battery systems. During the test, the fuel flow rate (46.3 g/hr on average) was adjusted manually using the burner valve to achieve a stable hot side temperature of 340±5 °C, the rated maximum sustained temperature for the TE module. At this fuel flow rate, an 8 oz (or 227 g) butane canister would perform for 4.9 hours. Such a flow rate represents about 25% of maximal fuel flow; at maximum flow one canister would last 1.2 hours with the burner tested.

## 3. Results

During the test, the device generated an average 4.6 W for 2.6 hours totaling 12 Wh of energy using 121.7 grams of butane, an amount possessing 1655 Wh of chemical energy. Extrapolating this energy production rate for the entirety of the available fuel establishes a 22.4 Wh estimate of total energy able to be generated. This result represents a system thermal efficiency of 0.7% and (assuming 60% heat loss through exhaust gasses) a device efficiency of 1.8%. The estimated burner thermal power was 629 W, with 252 W (40%) delivered to the TE module and 378 W (60%) lost to the exhaust. With the present setup we can use common heat sink airflow modeling to estimate a volumetric flow rate of 0.96 m³/min at 6 m/s and a total thermal resistance of 0.195 °C/W. The total device weight is 746 grams including 227 grams of butane fuel. This demonstrates a fuel mass fraction of 30%, slightly higher the stated figure (23%) for achieving parity with lithium battery technology with an existing high-performance (12% efficiency) TE module. With further development a system weight savings of 20% could be within reach, e.g. Polyether ether ketone (PEEK) fuel vessel, elimination of bayonet-Lindal valve adapter, and optimized heat sink design. Such a design could achieve an estimated fuel mass fraction of 36%, permitting the inclusion of ignition and fuel control components.

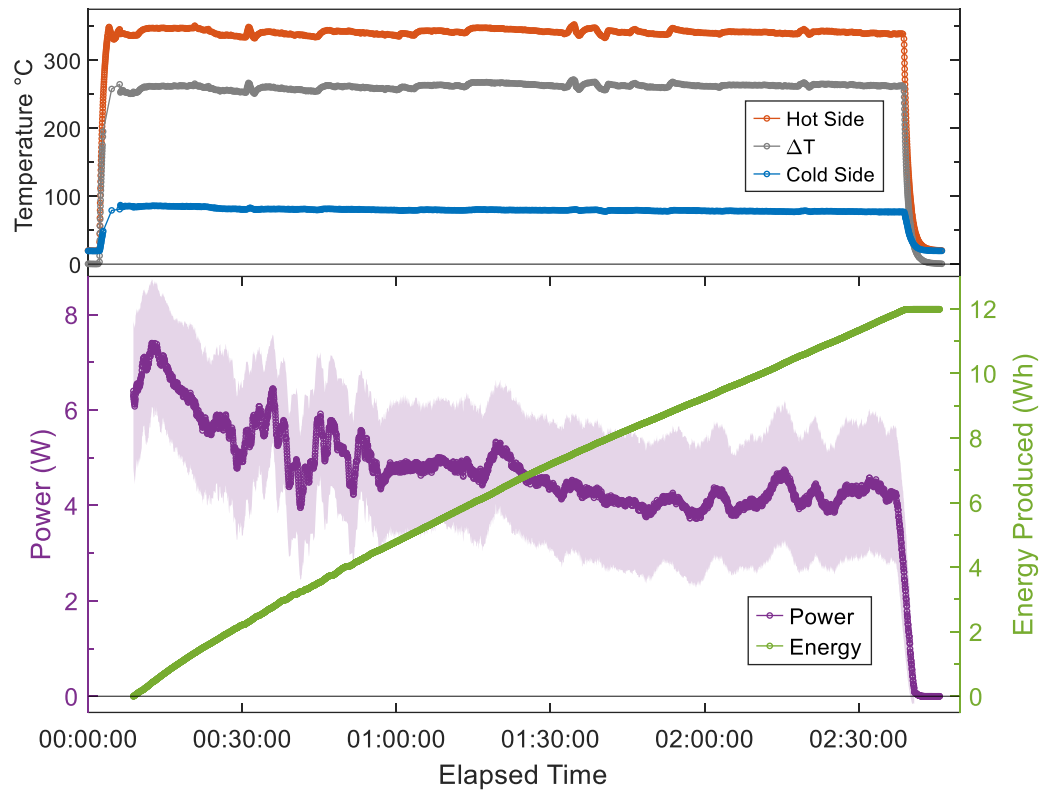

**Figure 4.** Results of a 2.6-hour long test of the prototype thermoelectric generator. The upper panel displays the temperature of the hot and cold sides of the TE module as well as the difference in temperature, $\Delta T$. Kinks in the temperature graph designate manual retrimming of the fuel flow to maintain 340±5 °C. The bottom panel shows power and total energy produced over the duration of the test. A 100-point (200 second) sliding window mean was applied to the power data for smoothing. The highly variable power produced is likely due to issues with the internal MPPT within the electronic load test device as voltages measured directly off the device were stable.

Overall, tests demonstrate the prototype generator (at 1.8% device efficiency) possesses specific energy and power of 30 Wh/kg and 6 W/kg, respectively. Ideally, specific energy and power increase proportionally with efficiency, therefore we might expect the following values with a 12% efficient TE module: 200 Wh/kg and 40 W/kg. With 20% weight savings the values become 250 Wh/kg and 50 W/kg, slightly beating battery specific energy, but possessing much lower specific power. Despite these weaknesses, important insights can be drawn from these results.

Firstly, this work was successful in achieving its novel objective: developing a lightweight physical prototype thermoelectric generator with the purpose of vetting a realistic fuel mass fraction. In this way, we have demonstrated a fuel mass fraction of 30%, exceeding the minimum 23% calculated for achieving parity with lithium battery specific energy. Unfortunately, estimates for TE-based specific power of 40 W/kg do not compete with the multi-thousand W/kg specific power of lithium batteries.

Secondly, the maximum efficiency of the TCS monTEG device is 5%, which is typical of Bismuth-Telluride based devices. The TCS device was chosen for budgetary reasons and was intended to serve as a placeholder device with results able to be extrapolated to higher performance Pb:TAGS (Te-Ag-Ge-Sb) modules with 2.4 times the efficiency (12%).

Thirdly, it is suspected that the maximum efficiency of the particular device (5%) was not achieved due to deformation of the hot side aluminum heat sink. The heat sink resists the bending moment of the screws along the fin axis, but is weak on the axis transverse to the fins. It is likely that weak bending resistance and high temperatures combined to deform the lapped surface of the heat sink along one axis. This issue likely explains the slight degradation of power production within a test as temperature induced creep would affect the system. A previous prototype heat sink with very thin fins displayed similar "doming" and poor thermal performance. There is a strong chance that a custom machined heat sink with reinforcements would cure this characteristic.

Lastly, there is a tradeoff relationship between specific power and specific energy. Increasing the specific energy amounts to increasing fuel mass, thereby decreasing specific power and vice versa. The prototype design represents a single point on the specific-energy-power landscape. Indeed, every application would have a different optimal set of parameters balancing specific energy and power.

The performance of the prototype device demonstrated that the overall design of the device achieves sufficient capability for generating power as a proof of concept and a foundation suitable for possible improvement. It also suggests that the fuel mass fraction achieved can appropriately serve as a valid contextual estimate.

*3.2. Endurance Modeling*

An endurance modeling script was developed in MATLAB that estimates the amount of time required to deplete the platform's power source at average power. The goal with this subtask of the work is to equitably compare performance of fully electric and fully butane systems, as well as any degree of hybridization in between. We parametrically sweep the fuel and battery energy quantities independently, probing any combination of hybrid operation, as shown in figure 5. The computation results in a 3D plot with the x and y axes representing inputs of energy in battery or fuel forms, while the z axis represents the estimated endurance. The model integrates TE module efficiency, battery/fuel energy density, fuel mass fraction, etc. The model also compensates mass dependent power requirement levels thereby providing a more accurate estimate for the deleterious effects of any overloading with batteries or fuel.

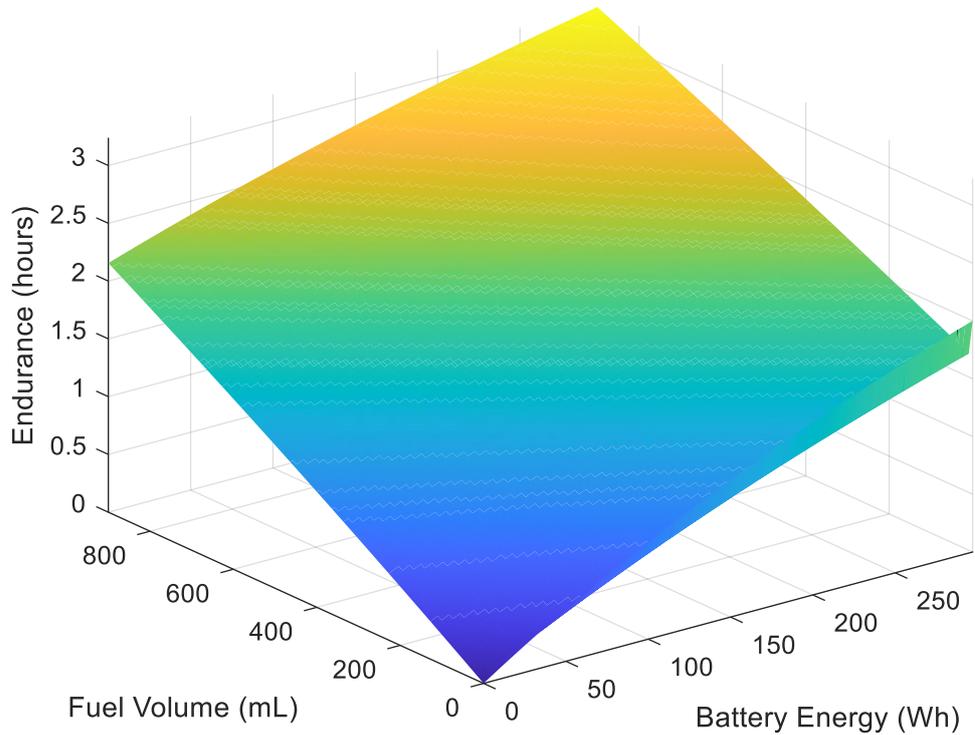

**Figure 5.** Modeling AeroVironment RQ-20 Puma endurance when supplied by battery and butane supplied energy as well as any combination. The sharp dip down close to the pure battery energy axis displays the "dead weight" penalty associated with TE generator use with low fuel quantity levels. Maximum butane-derived energy (accounting for efficiency losses) is comparable to maximum battery-derived energy (300 Wh each). The Puma represents the highest performing TE-powered drone within the sample due to its low specific power requirements. One can see that when using purely butane power the endurance is merely able to match the incumbent battery-powered method at 2 hours. Also, in practice, a limited battery system would be needed to absorb changes in drone power consumption as well as support reignition.

A sample of five unmanned system platforms were targeted for comparison and orientation in the use of the endurance modeling script. The sample included representatives from the following small unmanned platform classes: multicopter, fixed wing, fixed wing VTOL, and ground vehicle. Specifically, the sample included the Foster-Miller Talon, AeroVironment RQ-11 Raven, AeroVironment RQ-20 Puma, and Quantum Systems Trinity F90+. The empty mass, battery mass, stated endurance, and average required specific power (shown in Table 1) were gathered or calculated from available literature, with stated maximum endurance times reproduced in the model accordingly. Since this information is not always readily available, this sample represents systems in which these values could be obtained. Also, care must be taken in comparing platform specific power requirement and thermoelectric generator specific power. The former incorporates the airframe/chassis mass of the unmanned system, while the latter does not. Therefore, although the thermoelectric generator specific power figure of 40 W/kg for the prototype extrapolation may seem to qualify for most of the sample applications, it does not account for the airframe/chassis.

The maximum fuel volume quantity, of which endurance is highly dependent upon, was carefully chosen according to the criteria detailed in Appendix B.

**Table 1.** Characterization of relevant modeling inputs for each sample article as well as estimated device efficiencies, $\eta_{dev}$, modeled for comparable and double endurance performance.

| Platform Samples | Specific Power Requirement, W/kg | Maximum Fuel Quantity, mL | $\eta_{dev}$, Lithium Parity | $\eta_{dev}$, Twice Lithium Endurance |
|---|---|---|---|---|
| Talon | 3 | 4950 | 7.8% | 15.5% |
| Aurelia X6 | 90 | 796-1114 | 28% | 35% |
| Raven | 17 | 30-50 | 35% | 42.5% |
| Puma | 21 | 900-1023 | 10.5% | 17% |
| Trinity | 31 | 675-871 | 11.8% | 17.5% |

As shown with the prototype, parity can be achieved for the specific energy metric of butane powered thermoelectric systems only at the expense of specific power. This result is also borne out from results of modeling the AeroVironment RQ-20 Puma shown in figure 5. In fact, the low-power, moderately sized drones (Puma, Trinity, and the ground-based Talon) were the only members in our sample that could reach parity with lithium batteries. These results prompt the investigation of the degree to which the most salient metric, device efficiency, must increase in order for the model thermoelectric generator system to achieve comparable endurance durations. Required TE device efficiencies are shown in table 1. Note that these figures are a result of comparing platform endurance, not simply specific energy. In this way, these results account for practical complications such as the possession of sufficient specific power, incorporating volume and mass constraints as well as valid fuel mass fractions.

Oftentimes, the switch from incumbent technology to any alternative solution requires not only similar performance, but an improvement of two or three times in order to incentivize the transition as well as compensate for new technology adoption issues. Therefore, we list the required TE efficiency (and corresponding increase in maximum fuel quantity) for a doubling of platform endurance in the remaining column in Table 1.

## 4. Discussion

In this work, we have concentrated on comparing butane-fueled thermoelectric generation to the incumbent lithium-ion battery technology for application in small unmanned systems. The following is a list of additional possible power source alternatives: fuel cells, high voltage tethering, and inductive charging. Of the alternatives to lithium battery technology, thermoelectric and fuel cell solutions present themselves as the most suitable drop-in replacement for drone power sources, with thermoelectric solutions demonstrating its technology readiness advantage, although thermophotovoltaics are showing promising technology readiness levels and high efficiencies [10]. High voltage tethering could be an attractive alternative providing practically endless endurance, but sacrifices much of its mobility as well as swarming capability due to high voltage line tangling complications. Inductive charging (AKA power beaming), whether optical or microwave, is subject to similar advantages and limitations as tethering, albeit with a longer, more flexible range (more sophisticated electronics requirements as well). Inductive charging along powerlines is a strategy that may avoid these limitations, but would require a peace-time-like infrastructure operation environment typically not reliably assumed during battlefield operations.

For thermoelectric generation, alternative fuel sources were considered in addition to butane including solid fuel such as powdered magnesium, commonly used in Meal, Ready-to-Eat (MRE) heaters. MRE heaters are ubiquitous among armed forces ensuring availability and familiarity. They also possess impressive specific energies comparable to hydrocarbon fuel combustion despite relying on a slower oxidation process. However, due to their solid nature internal transport within a generator system would be impractical. Therefore, the fuel quantity would need to be equal to the area of the TE module in order to fully utilize its surface area. The capture of heat energy away from the TE module surface would be increasingly sparse as the volume of the fuel increases. The ease of internal transport is a crucial advantage of liquid or gaseous fuels, as isolated burners can be supplied with a fuel storage volume independent of the size of the TE module. For these reasons butane was pursued in this study.

Following the thorough characterization of the performance of butane-fueled thermoelectric generation, the following question presents itself: Given its performance profile, what applications would suit the advantages that butane-fueled thermoelectric generation possesses?

In light of the areas where lithium battery technology shines, TE systems would best be suited for applications which require high specific energy, very low specific power, portability, and long-term endurance (weeks to months), putting them closer in application to portable solar panels. But, of course, TE systems would not need direct solar access, therefore we may submit that they would best be suited for applications without direct sunlight, but with adequate ventilation. Possible applicable environments include austere locations with high amounts of cloud or canopy cover, within semi-permanent facilities, nighttime, or underground. Furthermore, these more static types of applications would lend themselves more flexible alternative fuel options e.g. MRE heaters, which could be easily swapped.

## 5. Conclusions

To conclude, we developed a proof-of-concept prototype butane-fueled thermoelectric generator able to produce 4.6 Watts with a fuel mass fraction of 30%, an estimated device efficiency of 1.8% and specific energy and power of 30 Wh/kg and 6 W/kg, respectively. These results can be extrapolated with an efficiency of 12% to 200 Wh/kg, and 40 W/kg if the highest performing TE module from MicroPower is used. Informed by these results, we conclude that thermoelectric generation can compete with lithium battery specific energy in some regimes, but is not competitive regarding specific power. We leverage the insights gained from the prototype to inform a MATLAB script developed to estimate drone endurance, inspecting a sample of five representative small unmanned systems including air- and ground-based platforms as well as VTOL and fixed wing configurations. We find that for some fixed wing applications with low specific power requirements thermoelectric generation can be comparable to lithium solutions. But more typical drone applications require around double the TE device efficiency in order to achieve parity with lithium solutions. Furthermore, increases in efficiency of three to four times would be required to double the most power-hungry drones' endurance.

To remark on ongoing thermoelectric materials research, it is difficult to overstate the importance of device maximum temperature for the drone endurance application. Endurance depends on efficiency, specific energy, and specific power, as is evident from this work. The fact that TE device power scales as $\propto \Delta T^2$ [4], and efficiency scales with $\Delta T$ [11], suggests that endurance would scale very favorably with $\Delta T$. Therefore, increasing the maximum operational device temperature likely becomes the most relevant metric to optimize.

Overall, the popularity and utility of small air- and ground-based drones are expected to continue increasing. The tension pitting the substantial utility of group 1 unmanned systems against their limited endurance, especially for multicopter drones, remains an obstacle for the foreseeable future. This limitation is so commonly recognized that many research works have implemented deep reinforcement learning approaches that specifically prioritize the minimization of energy consumption to conserve flight time when determining agents' actions [12–14]. These actions and policies reiterate conclusions that energy is a scarce resource for drones and the provision of more would alleviate control algorithm constraints and further augment drone utility.

**Supplementary Materials:** The following supporting information can be downloaded at:, Arduino Temperature Logging Script: TemperatureLogger.ino; Processing Temperature Log CSV Saving Script: LogTempCSV.pde; Temperature Data: SensorData2023-11-15_15_29_00.csv; Electronic Load Tester Power Data: 2023-11-15-18-32-20-EBC-A10H-1-1.csv; MATLAB Power and Temperature Data

Analysis and Graphing Script: DataAnalysisGraphing.m; MATLAB Drone Endurance Modeling: DroneEnduranceModeling.m.

**Author Contributions:** Conceptualization, M.W.; methodology, M.W.; software, M.W., A.R., and E.S.; validation, M.W., A.R., and B.W.; formal analysis, M.W.; investigation, M.W., A.R., and B.W.; resources, M.W., M.J.H., and A.R.; data curation, M.W. and B.W.; writing—original draft preparation, M.W.; writing—review and editing, M.W. and M.J.H.; visualization, M.W.; supervision, M.W. and M.J.H.; project administration, M.W.; funding acquisition, M.W. All authors have read and agreed to the published version of the manuscript.

**Funding:** This work was supported by the ARL:UT Independent Research and Development Program.

**Data Availability Statement:** All data is gathered in supplemental material.

**Conflicts of Interest:** The authors declare no conflicts of interest.

**Appendix A**

Table A1. Prototype component mass breakdown table.

| Component | Mass, g |
|---|---|
| Assembly without fuel canister | 413.0 |
| Full 8 oz butane canister | 333.4 |
| Cold side heat sink | 223.0 |
| Hot side heat sink | 53.0 |
| Bayonet-to-Lindal adapter | 36.7 |
| TE module | 26.2 |
| Ceramic insulation | 25.7 |
| Compression hardware[3] | 19.5 |
| Thermocouple circuit | 12.2 |
| Reignition module | 19.8 |
| Exhaust pipe | 15.6 |
| Transition duct | 70.0 |

**Appendix B**

The following section shows endurance modeling results for the sample of drones examined including sizing of drone components and comparisons with the respective established lithium battery components.

Table B1. Details of modeling input and output parameters. Battery figures are derived from drone specifications. Generator figures including mass, max fuel and combustion chamber (CC) volume, max fuel energy, fuel mass fraction, and thermoelectric side length are output results from endurance modeling script. TE surface side length is the side length of a square sufficient to cover the total surface area of TE module(s) for required power.

| Platform Samples | Battery Energy, Wh | Battery Mass, kg | Max Generator Mass, kg | Battery Volume, L | Generator Volume, L | Max Fuel Energy, Wh | Fuel Mass Fraction (at parity) | Volume Fraction | TE Surface Side Length (at parity), cm |
|---|---|---|---|---|---|---|---|---|---|
| Talon | 1350 | 9.5 | 6.0 | 5.0 | 5.0 | 1185 | 47% | 100% | 17 |
| Aurelia X6 | 710 | 5.0 | 4.7 | 2.6 | 1.3 | 691 | 9.6% | 49% | 21 |
| Raven | 32.5 | 0.23 | 0.23 | 0.12 | 0.04 | 32.5 | 7.3% | 37% | 4.9 |
| Puma | 297 | 2.1 | 2.1 | 1.5 | 0.94 | 280 | 23% | 85% | 15 |
| Trinity | 259 | 1.8 | 1.8 | 1.3 | 0.77 | 247 | 21% | 80% | 16 |

The volume fraction represents the volume of the fuel plus combustion chamber compared to the volume of the battery of the respective drone. The combustion chamber volume figure was estimated to be proportional to the power drawn with a proportionality constant (1.4 $\frac{mL}{W}$) established using the prototype. Due to the drone endurance being highly dependent on fuel quantity, specific criteria were required to keep modeling results realistic and relevant. We chose particular maximum fuel quantities according to the following rules: first, total generator mass (including fuel) could not exceed the original battery mass. Second, the total volume of the fuel and combustion chamber could not appreciably exceed the volume of the original battery. We do not include the volume of the cold side heat exchanger because it would be positioned exterior to the drone within outside airflow. One can see that a platform is typically either volume or mass constrained with the exception of the Aurelia X6 which is power constrained.

Output results of the endurance modeling are shown below in each figure for device efficiencies corresponding to lithium parity: (a) endurance vs battery energy vs fuel volume, (b) mass breakdown; and the same for twice lithium parity: (c) and (d).

---

[3] McMaster-Carr part descriptions and numbers: Titanium Socket Head Screw M4 x 0.70 mm Thread Size, 30mm Long 95435A976; 302 Stainless Steel Corrosion-Resistant Compression Springs 14.5mm Long, 9.6 mm OD, 6.4 MM ID 2006N234, Low-Strength Steel Coupling Nut Zinc-Plated, M4 x 0.7 mm Thread 93355A310

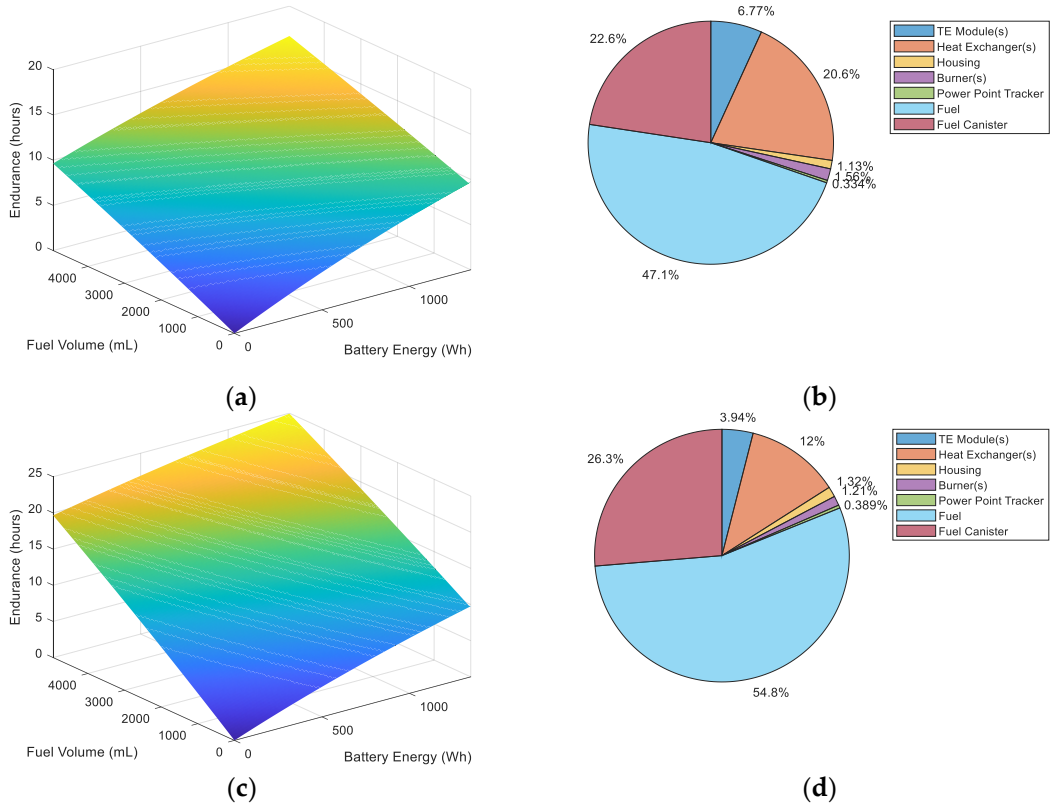

**Figure B1.** Modeling the Foster-Miller Talon at 7.8% device efficiency (**a**) Endurance; (**b**) Mass breakdown. At 15.5% device efficiency (**c**) Endurance; (**d**) Mass breakdown.

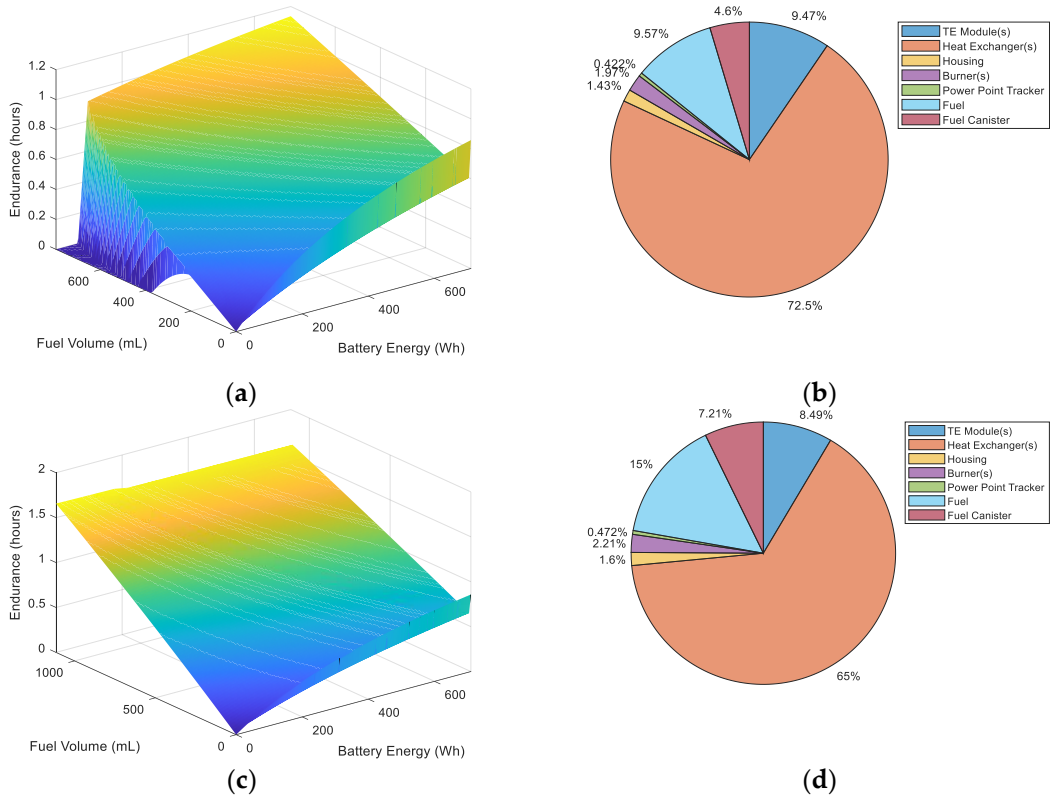

**Figure B2.** Modeling the Aurelia X6 at 28% device efficiency (**a**) Endurance (sharp drop-off at high fuel volume and low battery energy is due to insufficient power produced by TE module); (**b**) Mass breakdown. At 35% device efficiency (**c**) Endurance; (**d**) Mass breakdown.

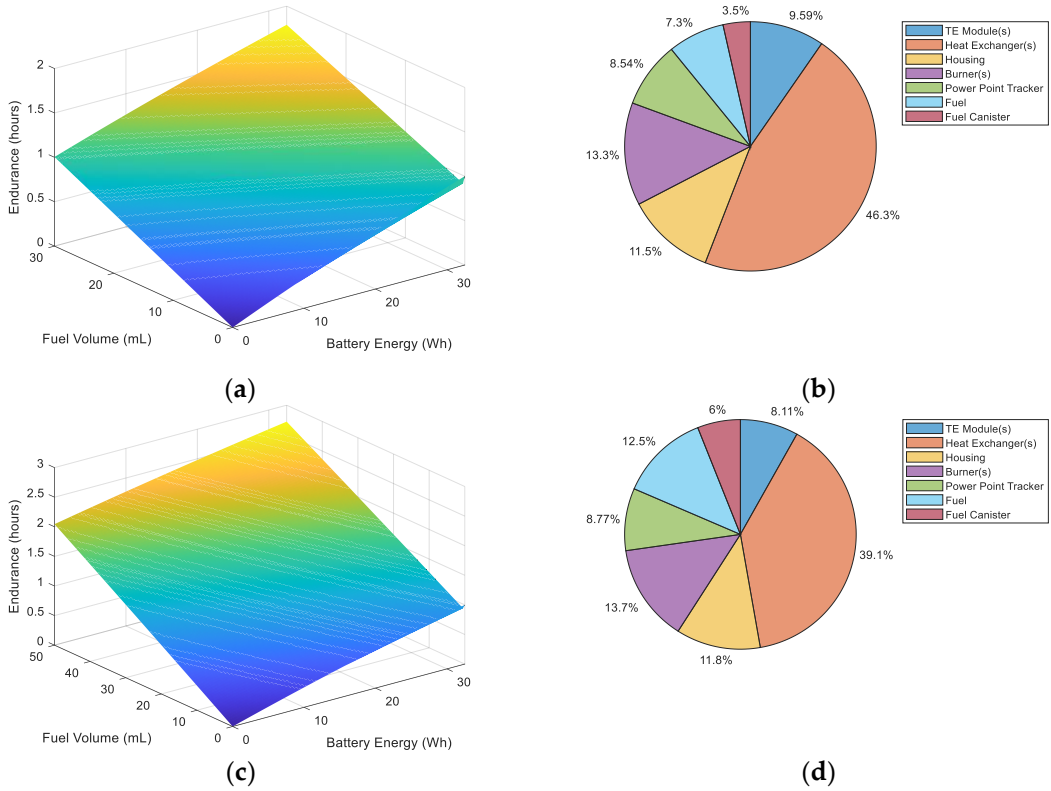

**Figure B3.** Modeling the AeroVironment RQ-11 Raven at 35% device efficiency (**a**) Endurance; (**b**) Mass breakdown. At 42.5% device efficiency (**c**) Endurance; (**d**) Mass breakdown.

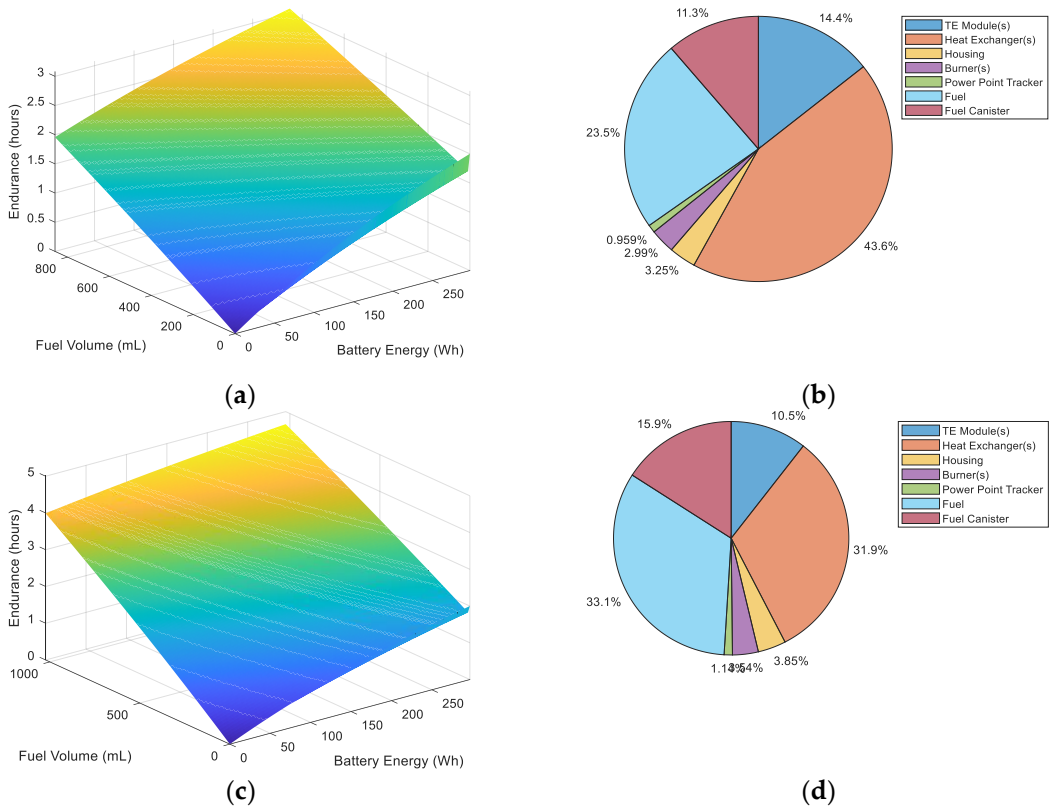

**Figure B4.** Modeling the AeroVironment RQ-20 Puma at 10.5% device efficiency (**a**) Endurance; (**b**) Mass breakdown. At 17% device efficiency (**c**) Endurance; (**d**) Mass breakdown.

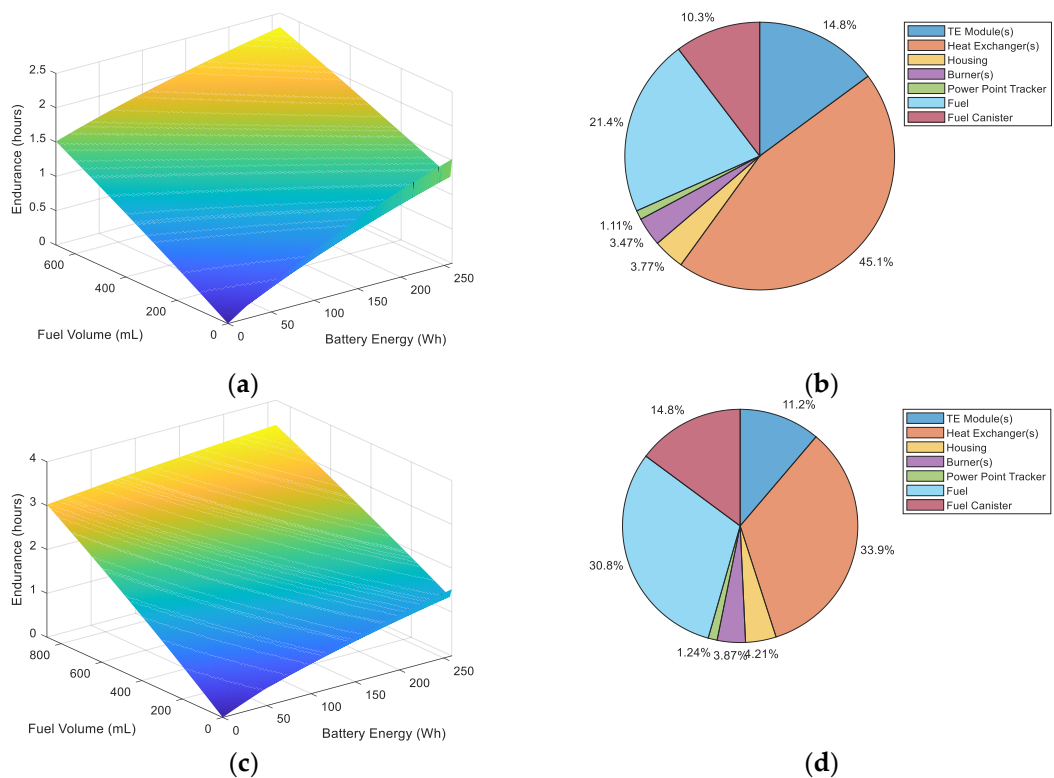

**Figure B5.** Modeling the Quantum Systems Trinity F90+ at 11.8% device efficiency (**a**) Endurance; (**b**) Mass breakdown. At 17.5% device efficiency (**c**) Endurance; (**d**) Mass breakdown.